\documentclass[10pt,final,journal,twocolumn]{IEEEtran}
\ifCLASSINFOpdf
\else
\fi
\usepackage{cite}
\usepackage[cmex10]{amsmath}
\usepackage{amsmath}
\usepackage{algorithm}
\usepackage{algorithmic}
\usepackage{stfloats}
\usepackage{multicol,multienum}
\usepackage{multirow}
\usepackage[dvips]{graphicx}
\usepackage{wrapfig}
\usepackage{color}
\usepackage{subfigure}
\begin{document}
%\title{A Dynamic User-Scheduling-based Hierarchical Power Control in Two-tier Femtocell Networks from the Perspective of Energy Efficiency}
\title{Stackelberg Game Approaches for Anti-jamming Defence in Wireless Networks}

\author{
       Luliang~Jia,~\IEEEmembership{Student Member,~IEEE,}
       %Fuqiang~Yao,
        Yuhua~Xu,~\IEEEmembership{Member,~IEEE,}
        Youming~Sun,~\IEEEmembership{Student Member,~IEEE,}
        Shuo~Feng,~\IEEEmembership{Student Member,~IEEE,}
        and~Alagan~Anpalagan,~\IEEEmembership{Senior Member,~IEEE}
        %and~Yonggang~Zhu
        % <-this % stops a space

%\thanks{This work was supported in part by the Natural Science Foundation for Distinguished Young Scholars of Jiangsu Province under Grant BK20160034, in part by the National Science Foundation of China under Grant 61631020, Grant 61671473, Grant 61401508, and Grant 61401505, in part by Jiangsu Provincial Natural Science Foundation of China Grant BK20130069, and Grant BK20151450, and in part by the Open Research Foundation of Science and Technology in Communication Networks Laboratory. Part of this paper had been submitted to the 2017 IEEE International Conference on Communication Technology (ICCT) (Corresponding author: Y. Sun.)}

\thanks{
L. Jia, Y.~Xu, Y.~Sun, and S.~Feng are with the College of Communication Engineering, Army Engineering University of PLA, Nanjing 210007, China. (e-mail:
jiallts@163.com; yuhuaenator@gmail.com; sunyouming10@163.com; fengs13@mcmaster.ca).}
%\thanks{
%F.~Yao is with  Nanjing Telecommunication
%Technology Institute, Nanjing 210007, China (e-mail: yfq2030@163.com).}
%\thanks{Y.~Sun is with National Digital Switching System Engineering $\&$ Technological Research Center, Zhengzhou 450001, China (e-mail: sunyouming10@163.com).}
%\thanks{S.~Feng is with the Cognitive Systems Laboratory, McMaster University, Hamilton, ON L8S 4L8, Canada (e-mail: fengs13@mcmaster.ca).}
\thanks{A.~Anpalagan is with the Department of Electrical and Computer Engineering, Ryerson University, Toronto, ON M5B 2K3, Canada (e-mail:alagan@ee.ryerson.ca).}
% <-this % stops a space
%\thanks{
%F.~Yao , Y. Niu, and Y. Zhu are with  Nanjing Telecommunication
%Technology Institute, Nanjing 210007, China (e-mail:yfq2030@163.com; niuyingtao78@hotmail.com; zhumaka1982@163.com).}% <-this % stops a space
}

\maketitle

\begin{abstract}
This article investigates the anti-jamming communications problem in wireless networks from a Stackelberg game perspective. By exploring and analyzing the inherent characteristics of the anti-jamming problem, we present and discuss some technical challenges and fundamental requirements to address them. To be specific, the adversarial characteristic, incomplete information constraints, dynamics, uncertainty, dense deployment, and heterogeneous feature bring technical challenges to anti-jamming communications in wireless networks. Then, for the purpose of improving system performance, four requirements for anti-jamming communications are presented and discussed. Following the advantages of the Stackelberg game model in anti-jamming field, we formulate an anti-jamming decision-making framework based on the Stackelberg game for anti-jamming defence in wireless networks. Moreover, two preliminary case studies are presented and discussed for better understanding of the anti-jamming Stackelberg game problem. Finally, some future research directions are also provided.
\end{abstract}

%\begin{IEEEkeywords}
%Distributed channel selection, exact potential game, stochastic learning, interference mitigation, anti-jamming, hypergraph.
%\end{IEEEkeywords}

% \fi
\IEEEpeerreviewmaketitle

\section{Introduction}

Due to the shared and broadcast nature, the wireless transmission is highly vulnerable to various attacks, such as spoofing attack, eavesdropping attack, data falsification attack, jamming attack, and so on. Therefore, the security issue becomes a critical attribute of wireless networks, and has attracted extensive attention in the past decade. In this paper, we focus on the jamming attack  \cite{existwork2,existworkG,existworkBal}, which is a serious threat to the security of wireless networks, and deteriorates the system performance significantly. The anti-jamming problem is an interesting and challenging topic in wireless transmission, and also a vital issue for the security of the spectrum availability. Various countermeasures have been proposed to fight against jamming attacks, and they can be broadly classified into two types: spread spectrum based techniques and resource allocation based techniques \cite{existwork4}. However, spread spectrum based techniques, such as Frequency Hopping Spread Spectrum (FHSS), Uncoordinated Frequency Hopping (UFH), and Random Codekey Selection using Codebook DSSS (RCSC DSSS)\cite{existworkBal}, require wide-band spectrum, and therefore, are regarded to be spectrally inefficient. It is of great importance to develop anti-jamming schemes that have efficient spectral efficiency, especially in scarce spectrum scenarios. Thus, it motivates the optimal resource allocation based techniques (i.e., game theory based methods [5]-[13]), which constitute another family of anti-jamming methods and try to obtain effective use of resources. Moreover, with the development of the cognitive radio technology and artificial intelligence, the jamming attacks with higher-level intelligence pose a great challenge to the existing defence mechanisms. Therefore, it is extremely important and timely to develop efficient and flexible anti-jamming schemes.

Game theory is a powerful mathematical tool to adequately model and analyze the mutual interactions among players. Among the game theoretical models, Stackelberg game, as an important hierarchical game, stands out to capture the sequential interactions among players, and to make a strategic decision-making in wireless networks. Considering the different attributes and hierarchical behaviors of the legitimate users and jammers, Stackelberg game is a promising method to analyze the hierarchical interaction process between the legitimate users and jammers, and it is, therefore, a suitable framework for the sequential decision-making of the anti-jamming defence in wireless networks. The advantages for exploiting Stackelberg game in anti-jamming field are given as follows:

\begin{itemize}
\item The legitimate users and the jammers belong to different identities, and they have different attributes. Stackelberg game can capture the interactions among players with different attributes. Moreover, the legitimate users need to detect the jammers' actions to combat jammers, or the jammers are able to learn users' strategies to improve the jamming efficiency. Thus, the anti-jamming problem contains a natural hierarchical characteristic, and Stackelberg game is a suitable framework to capture and analyze the hierarchical interactions between the legitimate users and jammers.
\item The competitive interactions exist between the legitimate users and jammers (upper-level players and lower-level players), and the mutual competitive interactions also exist among the legitimate users (the players with same level) in dense wireless networks. Fortunately, the Stackelberg game model can simultaneously capture the competition at different levels.
\end{itemize}

Stackelberg game model has attracted growing attention in anti-jamming field recently, and some Stackelberg game solutions have been presented for anti-jamming defence in wireless communications. In \cite{existwork6,existwork7,existwork9,existwork10,existworkadd}, the Stackelberg game framework was employed to model and analyze the transmitting-jamming problem, and the anti-jamming power control game was investigated in wireless networks. In \cite{existwork11}, the cooperative transmission game was studied, and the equilibrium solution was obtained. In \cite{existwork12}, the authors formulated an attacker-defender Stackelberg game between a jammer and a target node, and the timing channel was exploited. In \cite{existwork13}, a secure offloading game was formulated, and the Stackelberg equilibrium was derived. In \cite{existwork14}, we investigated the anti-jamming channel selection problem in an adversarial environment, and proposed a hierarchical learning approach to obtain the desirable solutions.

%In the literature, the anti-jamming Stackelberg game model has been investigated, and some preliminary anti-jamming Stackelberg game solutions have been studied

Note that a survey on the jamming and anti-jamming techniques in wireless networks can be found in \cite{existworkG}. It investigated several types of jammers from the jamming point-of-view and summarized the existing anti-jamming techniques from the perspective of the system security. In this paper, however, we investigate the anti-jamming defence problem in wireless networks from a Stackelberg game perspective, and we mainly focus on exploring and analyzing the inherent features, fundamental requirements and technical challenges in anti-jamming communications research. The main contributions of this article are given as follows:
\begin{itemize}
\item We analyze and discuss the fundamental requirements and technical challenges according to the inherent characteristics of the anti-jamming defence problem in wireless networks.
\item We outline some advantages of the Stackelberg game for the anti-jamming defence in wireless networks, and formulate an anti-jamming decision-making framework based on the Stackelberg game.
\item We provide the future research directions of the anti-jamming defence in wireless networks.
\end{itemize}

 %analyze some novel application scenarios of the anti-jamming Stackelberg game

The rest of the article is organized as follows. In Section II, we discuss and analyze the technical challenges and fundamental requirements of anti-jamming defence in wireless networks. In Section III, the Stackelberg game-theoretic model is investigated, and an anti-jamming decision-making framework based on Stackelberg game is established. In Section IV, two preliminary case studies are given, and future research directions are discussed in Section V. Finally, concluding remarks are presented in Section VI.

% To the best of our knowledge,  there are few researches on dynamically hierarchical joint user-scheduling and power control in tiered femtocell networks considering energy consumption.

\section{Challenges and requirements for anti-jamming communications in wireless networks}

\begin{figure*}[!t]
\centering
\includegraphics[width=0.7\linewidth]{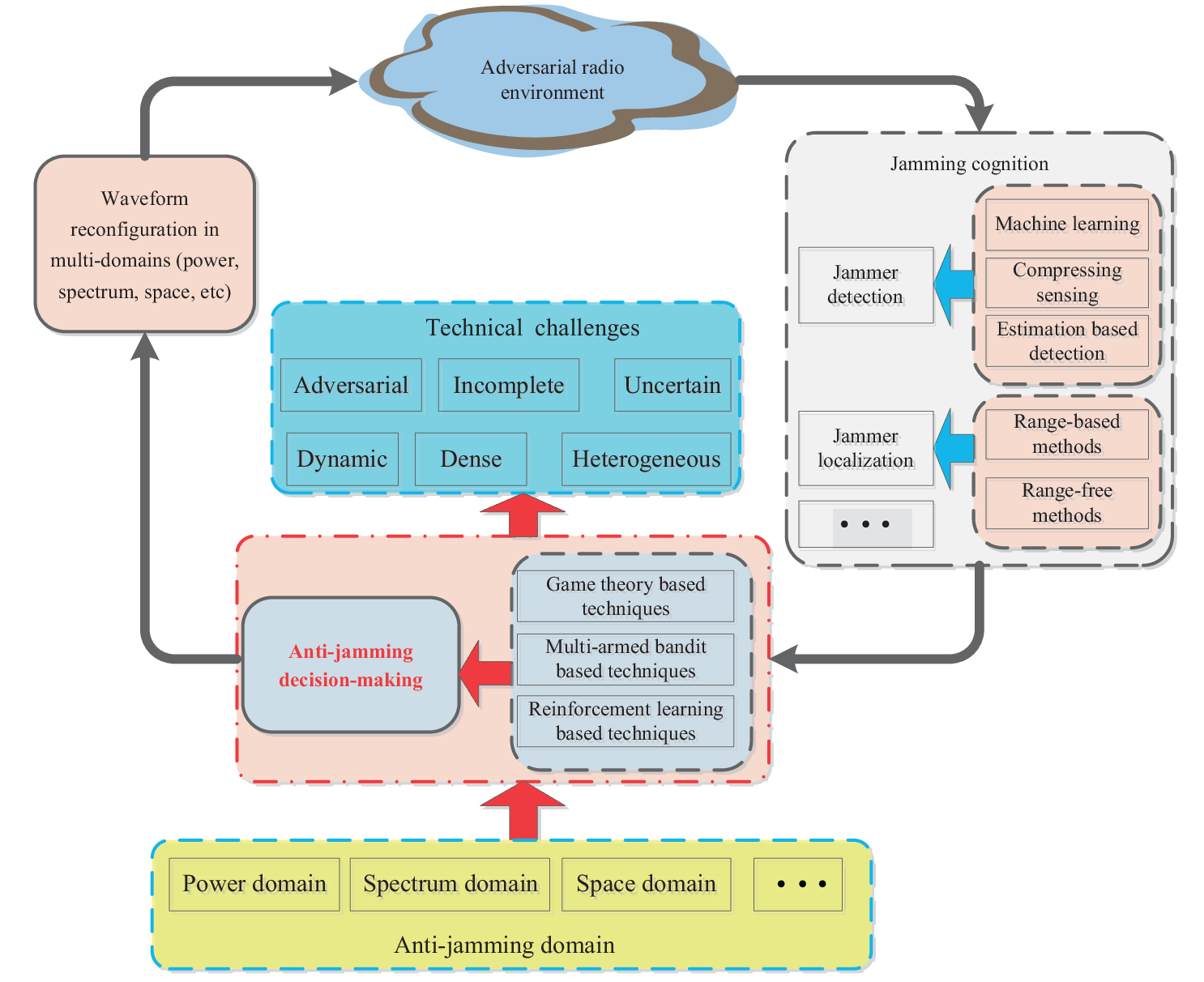}
\caption{Anti-jamming communications cycle in wireless networks.}
\label{Fig1}
\end{figure*}

From the perspective of the engineering, analogous to the cognition cycle, we formulate an anti-jamming communications cycle to describe the anti-jamming operations, which is the backbone for the anti-jamming communications. As shown in Fig.~\ref{Fig1}, the anti-jamming communications cycle mainly consists of the following steps: jamming cognition, anti-jamming decision-making, and  waveform reconfiguration.

\textbf{Jamming cognition:} The anti-jamming communications cycle begins with the jamming cognition, and the cognitive radio technologies can be employed to acquire useful information. It can perceive the jamming activities and complicated surrounding environment. Specifically, it should include jammer detection, jammer localization, and so on. For jammer detection, many techniques have been proposed, such as machine learning, compressing sensing, and estimation based detection method. The existing detection approaches were proposed for specific network environment and jamming models. However, it is a challenging task to differentiate a jamming scenario for the legitimate users due to the non-cooperative relationship between the legitimate users and jammers. The jammer localization is another important aspect, and various methods were provided, such as range-based methods (e.g., Pinpoint, WiSlow, and CrowdLoc) and range-free methods (e.g., Centriod Localization, Double Circle Localization, and Triple Circles Localization). Due to the adversarial characteristic and environmental factor, the inaccurate measurement and data insufficiencies are the main challenges in jammer localization. Note that the jamming cognition is not limited to sensing the surrounding environment and obtaining useful information, it can also realize the high-level intelligence in the near future, such as knowledge discovery and jamming prediction.

\textbf{Anti-jamming decision-making:} Based on the jamming cognition, the anti-jamming decision-making is performed in adversarial environments, and the optimal anti-jamming strategy (e.g., transmission power, channel) is chosen. It is the core technology of the anti-jamming communications, and a challenging task is to make an effective anti-jamming decision-making in wireless networks, especially in scarce resource scenarios and smart jammer scenarios. Due to the features of the wireless network and adversarial characteristic, the anti-jamming decision-making problem faces some challenges, which will be presented later. For the anti-jamming decision-making, various techniques were proposed in existing literatures, such as game theory based techniques, multi-armed bandit based techniques, reinforcement learning based techniques, and so on. Among these methods, game theory based techniques, especially the Stackelberg game based techniques, can adequately model the interactions between the legitimate user and jammer. In this paper, we mainly analyze the Stackelberg game based anti-jamming methods.

\textbf{Waveform reconfiguration:} The waveform reconfiguration can be accomplished in multi-domains, such as power domain, spectrum domain and space domain, to implement the anti-jamming communications. It is the physical measure to achieve the reliable transmission in jammed wireless networks. In power domain, we can fight against the jamming attack by adjusting the transmitting power. However, the increase of the transmitting power may deteriorate the linearity of power amplifier, and it will affect the performance of some modulation schemes that are sensitive to the linearity of power amplifier. In spectrum domain, the channel switching is adopted to cope with the jamming attack. Note that the channel switching may bring the performance loss, since the reconstruction of the communication link needs the settling time for the radio frequency (RF) devices. Moreover, different frequencies will have different propagation characteristics, and it may bring the difference in signal processing.

%For the considered scenario, it will have many potential application scenarios, such as wireless mesh networks and unmanned aerial vehicles networks.
In this article, we mainly focus on the anti-jamming decision-making process, which is the critical phase for anti-jamming communications in wireless networks. In the following, we aim to explore the inherent features of the anti-jamming communications, discuss and analyze the technical challenges and fundamental requirements for anti-jamming defence in wireless networks. In the following, we mainly consider the anti-jamming defence problem in power domain and spectrum domain.

\subsection{Discussion on technical challenges}
%%%%%%%%%%%%%%%%%%%%%%%%%%%%%%%%%%%%%%%%%%%%%%%%%%%%%%
For anti-jamming communications in wireless networks, it pursues the reliable transmission in the presence of the external malicious jammers. In this section, by exploring the characteristics of wireless networks \cite{existworkxu} and specificities in jammed networks, we discuss the technical challenges of the anti-jamming problem.

(1) \textbf{Adversarial:} In jammed wireless networks, the legitimate users always aim to maximize their utilities, and pursue reliable transmission, whereas the malicious jammers try to purposely deteriorate the intended transmission, and minimize the utility of the legitimate user. Therefore, the legitimate users and jammers work in an adversarial and non-cooperative environment.

Different from the general wireless networks, the adversarial characteristic is an important challenge in jammed networks. It is difficult to acquire the complete information of the opponent due to the adversarial characteristic, and the incomplete information constraints are common in anti-jamming decision-making. To cope with the incomplete information constraints, the Bayesian game framework and learning technologies (i.e., stochastic learning automata) are effective methods. In a Bayesian game framework, the utility function is defined over statistics, such as taking expectation, to describe the incomplete information constraints, and only the distribution information is needed. The learning technology is another candidate to deal with the incomplete information constraints. It can obtain the desirable solutions by trial-and-error exploration and learning from historical information.

(2) \textbf{Incomplete:} Due to the adversarial feature between the legitimate users and jammers, it is difficult to acquire complete information of the opponents. Furthermore, considering the limitation of the hardware and resource consumption, only partial information of the environment is available for both the users and jammers. Various forms of incomplete information constraints can be present, such as incomplete information of the channel gain \cite{existwork10} and the user type \cite{existworksa}.

(3) \textbf{Uncertain:} Due to the limitation of the capability of the signal processing and hardware facilities, it is difficult to observe the perfect information, and the observation error is common in anti-jamming field.

(4) \textbf{Dynamic:} The environmental information is variable over the course of time. For example, the channel state information is dynamic, and the traffic demands are time-varying due to their specific requirements. Furthermore, the position of the jammers may be random as well.

(5) \textbf{Dense:} In the coming future, dense networks are common, such as small cell networks and D2D networks. It will be challenging and interesting to investigate the anti-jamming problem in dense wireless networks, where both the mutual interference among users and external malicious jamming need to be simultaneously considered. Specifically, external jamming is a crucial factor that leads to the performance degradation, and the mutual interference among users is another factor that significantly restricts the network performance.

\begin{figure*}[!t]
\subfigure[An illustration of the dense wireless networks.]{
\label{fig:mini:subfig:a} %% label for first subfigure
\begin{minipage}[b]{0.5\textwidth}
\centering
\includegraphics[width=2.6in]{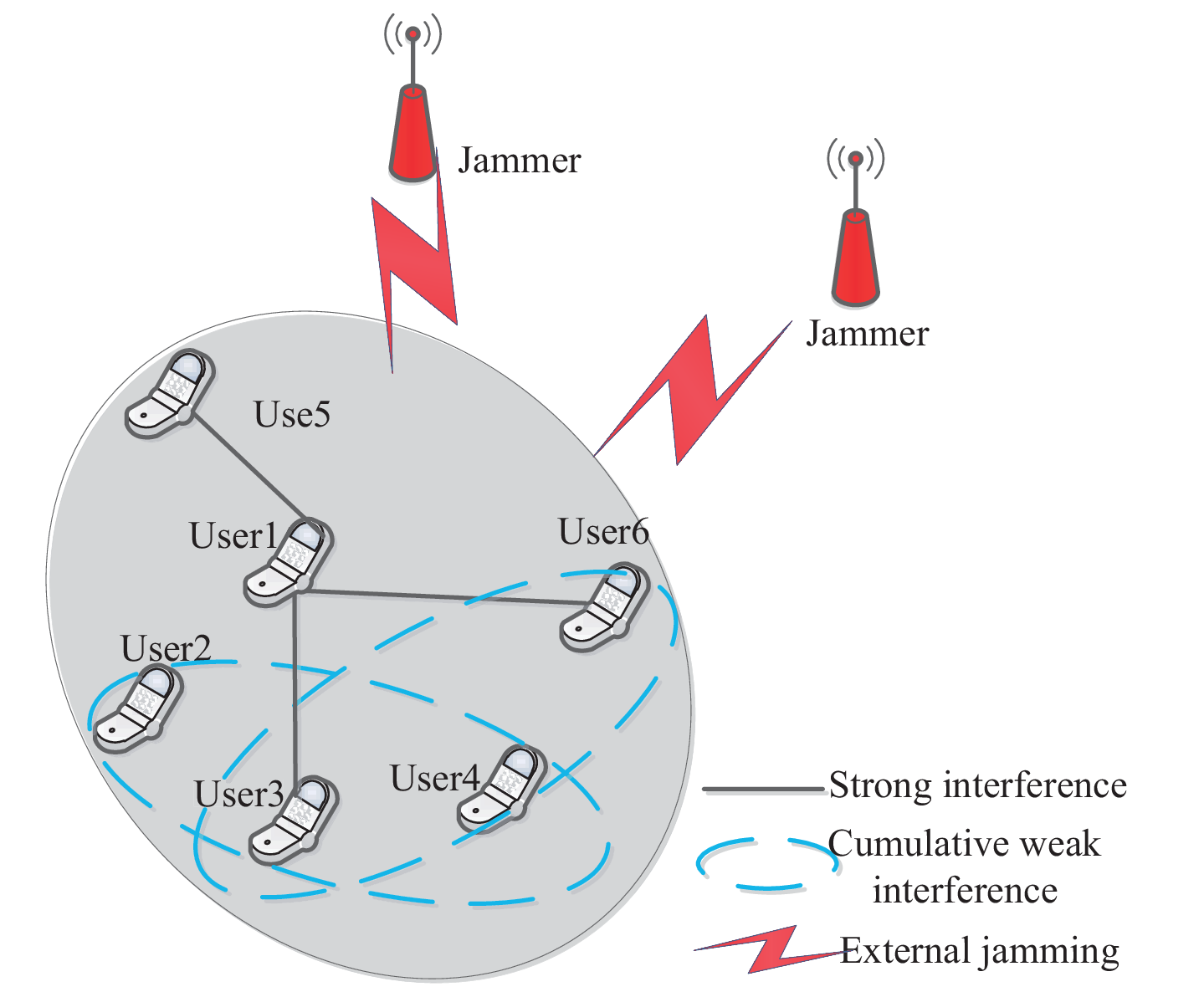}
\end{minipage}}%
\subfigure[An example of the heterogeneous networks.]{
\label{fig:mini:subfig:b} %% label for second subfigure
\begin{minipage}[b]{0.5\textwidth}
\centering
\includegraphics[width=2.5in]{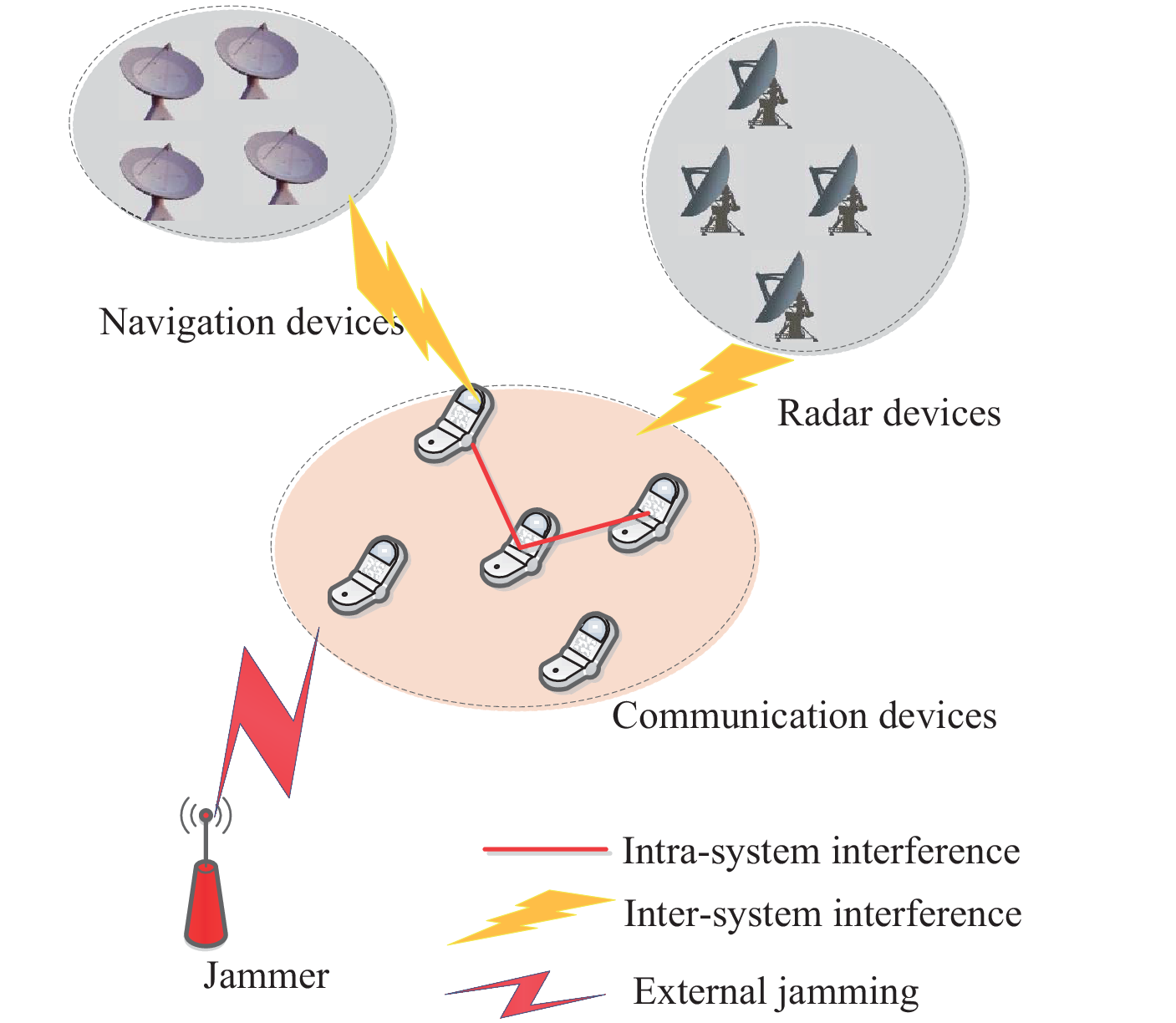}
\end{minipage}}
\caption{An illustration of the wireless networks}
\label{networks} %% label for entire figure
\end{figure*}

For the anti-jamming problem in dense networks, firstly, it is of paramount importance to fight against the external malicious jamming. Secondly, it is essential to consider the interference mitigation problem. In our prior work \cite{existwork14}, a preliminary anti-jamming scheme in dense wireless networks had been investigated, and a hierarchical learning framework was formulated. It is noted that the hypergraph model is a good candidate to explicitly describe the interference relation. As can be seen from Fig.~\ref{networks}(a), the hypergraph interference model can adequately capture the strong interference relation and cumulative weak interference relation. The dash lines in Fig.~\ref{networks}(a) represent the strong interference relation, and the circles denote the cumulative weak interference relation. It will be promising to incorporate hypergraph interference model into the anti-jamming game in dense wireless networks.

%\begin{figure}[!t]
%\centering
%\includegraphics[width=0.9\linewidth]{Fig2}
%\caption{An illustration of the dense wireless networks.}
%\label{Fig2}
%\end{figure}
%
%
%
%
%
%\begin{figure}[!t]
%\centering
%\includegraphics[width=0.9\linewidth]{Fig3}
%\caption{An example of the future heterogeneous networks.}
%\label{Fig3}
%\end{figure}

%%%%%%%%%ͨÀ¸Í¼Æ¬ÅÅ°æ%%%%%%
%%%%%%%%%%%%%%%%%%%%%%%%%%%
%%%%%%%%%ͨÀ¸Í¼Æ¬ÅÅ°æ½áÊø%%%%%%

(6) \textbf{Heterogeneous:} In future wireless networks, as indicated in Fig.~\ref{networks}(b), different types of devices will co-exist in a region, such as communication devices, navigation devices, radar devices, and so on. It will lead to a severe conflict and complicated environment. The intra-system interference, inter-system interference, and external malicious jamming need to be jointly considered.

In heterogeneous wireless networks, there exist diverse devices. Meanwhile, different devices belong to different systems (e.g., communication, navigation, and radar), and have different service demands (e.g., voice, image, and data). Different systems and services may have different anti-jamming requirements. Moreover, due to the dense deployment of diverse devices, the severe intra-system interference and inter-system interference bring new challenges. Therefore, it will be extremely challenging to investigate the anti-jamming problem in heterogeneous wireless networks, and we need to jointly cope with the intra-system interference, inter-system interference, and external malicious jamming.

\subsection{Discussion on fundamental requirements}
Considering the inherent characteristics, we list some fundamental requirements of the anti-jamming defence  problem in wireless communications, which mainly include reliability, robustness, scalability, and heterogeneity.

Firstly, it is necessary to obtain reliable transmission in wireless networks, and design effective anti-jamming approaches, especially in scarce resource scenarios and smart jammer scenarios.

Secondly, it should be robust to cope with different kinds of dynamics and uncertain factors in anti-jamming field. As discussed before, the channel state information is partially available. Moreover, the traffic demands may be variable.

Thirdly, as stated before, the user devices will be densely deployed in future wireless networks, and both mutual interference and external jamming will affect system performance. Therefore, the designed anti-jamming defence schemes should have the ability to be extended to dense scenarios, and address the mutual interference among users and malicious jamming simultaneously.

Finally, various heterogeneous devices will coexist in future wireless networks. It brings new challenges to the traditional anti-jamming solutions in homogeneous networks, and they cannot be directly employed. Therefore, it is essential to develop novel anti-jamming schemes in heterogeneous wireless networks.

\section{Anti-jamming Stackelberg game in wireless networks}

The Stackelberg game is regarded as an extension of the non-cooperative game, and it is a powerful mathematical tool that can be used to model the hierarchical interactions among players in a sequential manner. In a Stackelberg game, there are two types of players: leaders and followers. The leaders have the privilege over the followers and take actions first. Then, the followers make decisions based on the leaders' announced strategy. For a Stackelberg game model, the leaders and followers have their respective utility functions, and the most common solution is the Stackelberg Equilibrium (SE), which means no player can achieve higher utility by deviating unilaterally.

Stackelberg game model has been extensively applied to wireless networks recently, and it has been a powerful tool to make the hierarchical decision-making in wireless networks, such as offloading mechanism, enhancing secrecy rate, dynamic spectrum access in heterogeneous networks, wireless service provider selection, and power control in femtocell networks. In anti-jamming field, it is an appealing tool to adequately model and analyze the hierarchical interactions between the legitimate users and external jammers, and make a strategic decision-making in an adversarial environment.

To deal with the technical challenges of the anti-jamming issue in wireless networks, an anti-jamming decision-making framework based on Stackelberg game is formulated to combat the jammers. As shown in Fig.~\ref{Fig4}, it mainly consists of two steps:

\begin{figure}[!t]
\centering
\includegraphics[width=0.9\linewidth]{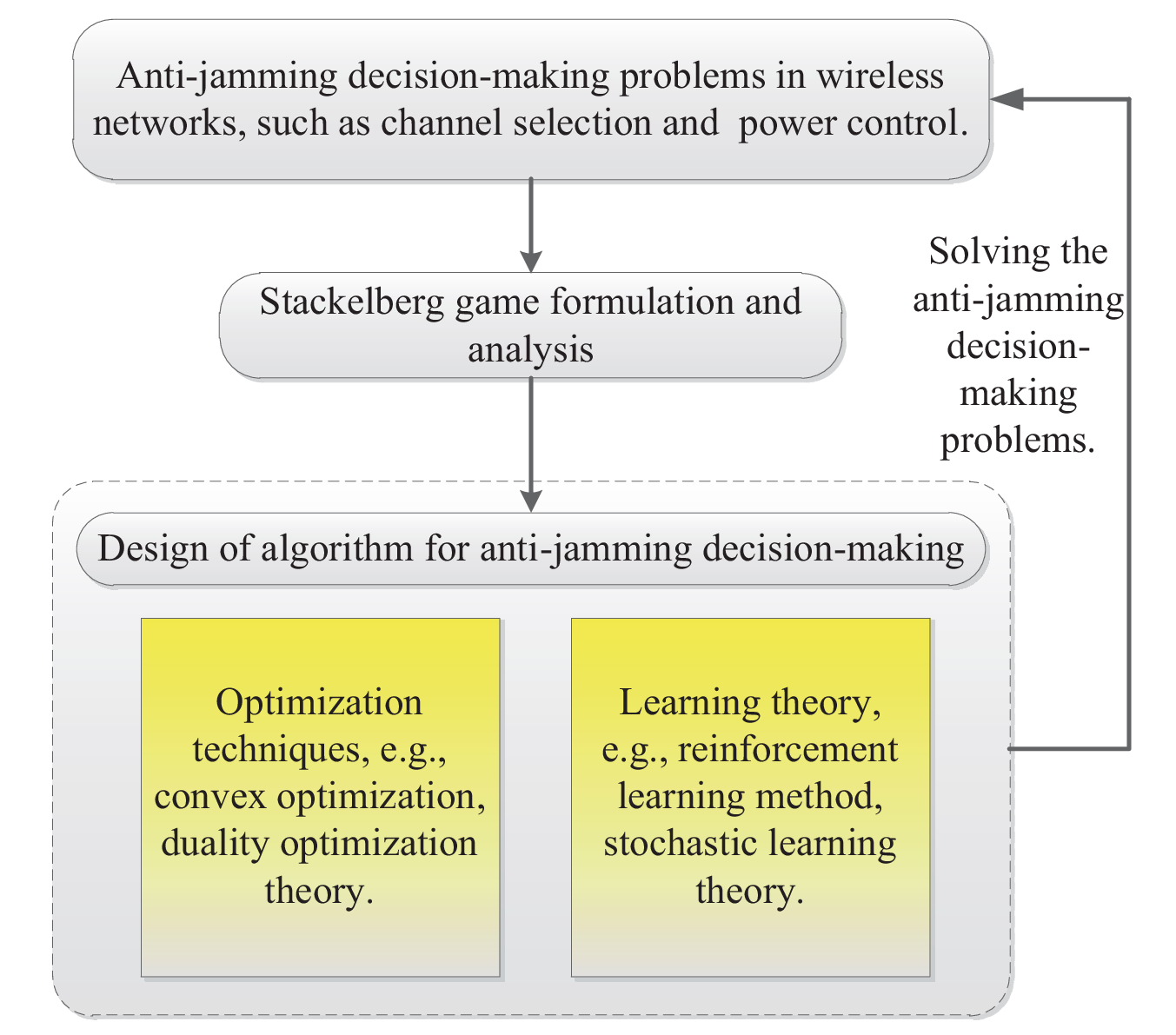}
\caption{The proposed anti-jamming decision-making framework in wireless networks.}
\label{Fig4}
\end{figure}

\begin{itemize}
\item Stackelberg game formulation and analysis.
\item Design of algorithms for anti-jamming decision-making.
\end{itemize}

\textbf{Stackelberg game formulation and analysis:} Firstly, it is necessary to recognize the identities and their available actions of the players, and design proper utility functions accordingly. Secondly, for an anti-jamming Stackelberg game, we need to answer who act as the leaders and who are followers, the answer for which is that it depends on specific research scenarios. For example, in \cite{existwork6}, \cite{existwork7}, \cite{existwork10}, and \cite{existworkadd}, it is assumed that the jammer is able to learn the legitimate user's strategies, and the legitimate user take actions first. The legitimate user is the leader, and the jammer is assumed to be the follower. In \cite{existwork9}, two scenarios (jammer as the leader and user as the leader) are both considered. In \cite{existwork14}, the users need to detect the jammer's actions before the channel selection, and it is assumed that the jammer acts as the leader and the users are followers. Since the utility function has a great impact on the performance of the game model, it is thus important to define proper utility functions. For example, the player can obtain the unique optimal strategy for the power control problem if the utility function is convex \cite{existwork10}. For the channel selection problem, the properties of the game model depend on the utility function, such as the exact potential game, which satisfies that the variation of the potential function is the same as the variation of the utility function by any player's unilateral deviation. The exact potential game has several attractive properties (i.e., admitting at least one pure strategy Nash equilibrium), and it can be incorporated into the follower sub-game \cite{existwork14}.

 Mathematically, an anti-jamming Stackelberg game can be formulated as ${\cal G} = \left\{ {{\cal N}{_u},{{\cal N}_j},{{\cal S}_u},{\cal S}{_j},{\mu _s},{\mu _j}} \right\}$, where ${\cal N}{_u}$ and ${{\cal N}_j}$, ${{\cal S}_u}$ and ${\cal S}{_j}$, ${\mu _s}$ and ${\mu _j}$ are player set, strategy space, and utility function of the legitimate users and jammers, respectively. In a Stackelberg game model, there exists a hierarchical competition between the upper-level players (leaders) and lower-level players (followers). Moreover, the mutual competition also takes place among the players within the same level. Fortunately, the Stackelberg game can adequately model and analyze the competition at different levels.

The Bayesian Stackelberg game is a good framework to cope with the uncertainties due to the uncertain and incomplete constraints. For the dense deployment scenario, Stackelberg game is a suitable tool, which can analyze the mutual competition among legitimate users and the competition between the legitimate users and external jammers simultaneously. For a future heterogeneous network, intra-system interference, inter-system interference, and external malicious jamming co-exist, and the multi-level Stackelberg game may be a good candidate for the anti-jamming problem in heterogeneous wireless networks.

\textbf{Design of algorithms for anti-jamming decision-making:} For the anti-jamming Stackelberg game problem, it is also a hierarchical optimization problem, and can be transformed into sequential sub-problems. To achieve this, a backward induction method is a common approach, and it has been employed to analyze the Stackelberg game and obtain the game-theoretic solutions. The follower sub-game is firstly investigated. Then, the leader sub-game is considered.

For the continuous problem, such as the continuous power control \cite{existwork6,existwork7,existwork9,existwork10,existwork11}, the SE solution is obtained based on the convex optimization techniques, such as Lagrange duality optimization theory, and Karush-Kuhn-Tucker (KKT) conditions. However, for the discrete problem, such as the anti-jamming channel selection problem \cite{existwork14}, it is intractable to adopt the traditional convex optimization techniques, and we need a new optimization framework and thus resort to learning theory, which achieves the desirable solutions through repeated trial-and-error exploration with a random environment. Some learning algorithms can be found in previous works, such as reinforcement learning, stochastic learning automata, and spatial adaptive play \cite{existworkxu}. The learning technology can cope with uncertain, dynamic, and incomplete information constraints, whereas game theory can adequately model and analyze the mutual interactions among users. Therefore, it is promising to incorporate the learning technologies into game theory. However, when the learning algorithms are incorporated into game theory, the challenge is to prove the convergence of the learning algorithms, which is application-dependent and will significantly differ for various applications.

%%%%%%%%%ͨÀ¸Í¼Æ¬ÅÅ°æ%%%%%%
%%%%%%%%%%%%%%%%%%%%%%%%%%%
\begin{figure*}[!t]
\subfigure[The process of the anti-jamming power control game.]{
\label{fig:mini:subfig:a} %% label for first subfigure
\begin{minipage}[b]{0.5\textwidth}
\centering
\includegraphics[width=2.2in]{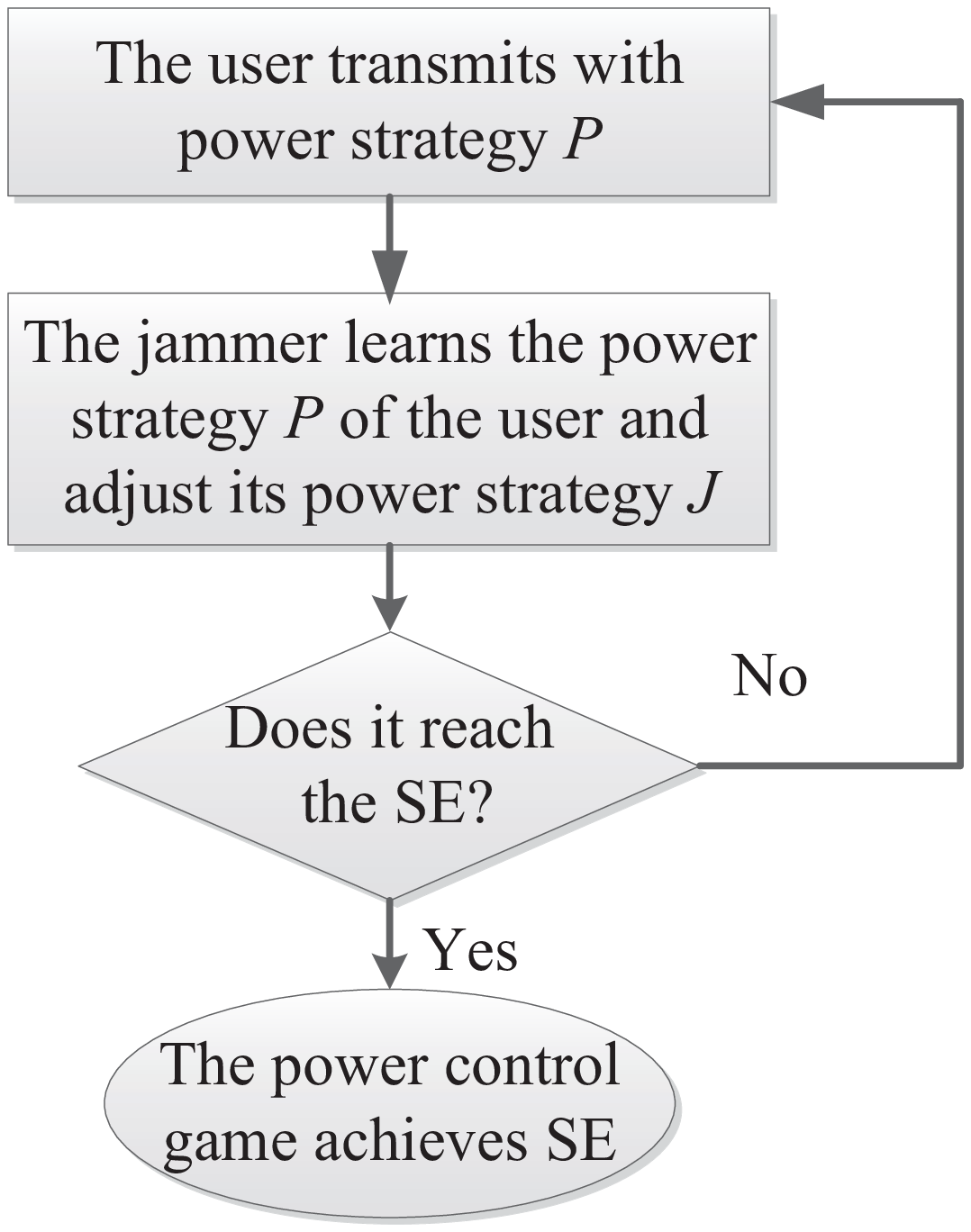}
\end{minipage}}%
\subfigure[The process of the anti-jamming channel selection game.]{
\label{fig:mini:subfig:b} %% label for second subfigure
\begin{minipage}[b]{0.5\textwidth}
\centering
\includegraphics[width=3.7in]{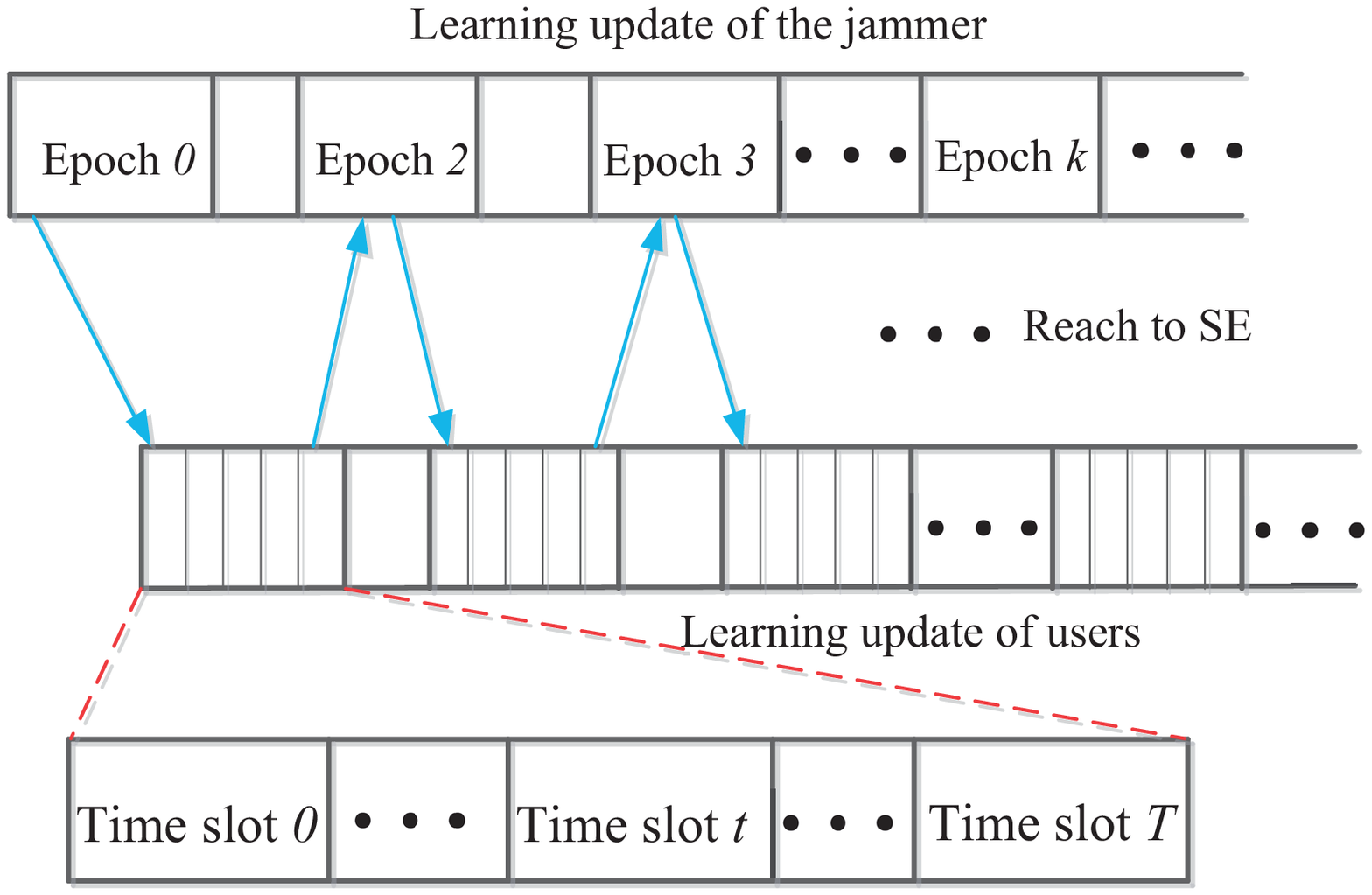}
\end{minipage}}
\caption{The implementation process of the anti-jamming Stackelberg game.}
\label{process} %% label for entire figure
\end{figure*}
%%%%%%%%%ͨÀ¸Í¼Æ¬ÅÅ°æ½áÊø%%%%%%

 \section{Case studies}
In this section, we present two case studies of anti-jamming Stackelberg game approaches to motivate the readers and appreciate the practicality of such approaches.

\subsection{Power control game with incomplete information}
Most existing anti-jamming Stackelberg power game approaches studied the scenarios with complete information. However, the scenarios with incomplete information are more general and practical in anti-jamming field. In \cite{existwork10}, we investigated the anti-jamming power control problem with incomplete information constraints in wireless communications. A system model with one user (transmitter-receiver pair) and one jammer is considered, and a Bayesian anti-jamming Stackelberg game is proposed, which can cope with the adversarial characteristic, incomplete information constraints. To capture the incomplete information constraints, the utility function is defined over statistics, such as taking expectation. Based on the duality optimization theory, the SE solution is derived. Moreover, the impact of the jammer's observation error $\varepsilon$ is analyzed. The utility function of the user is defined as the signal to interference plus noise ratio considering the transmission cost. The jammer's utility function is also defined as a function of the signal to interference plus noise ratio and the transmission cost. In the considered scenario, the user aims to maximize its utility, and takes actions first. The jammer can learn the user's transmission strategies, and play its best response strategy based on the user's announced strategy. The SE solution is obtained by the backward induction method, and the follower sub-game is investigated at first. To better understand the process of the power control game, a flow chart is shown in Fig.~\ref{process}(a). In this section, the transmission rate can be defined as:
\begin{equation}
R = \log (1 + \delta),
\end{equation}
where $\delta$ denotes the signal to interference plus noise ratio.

\begin{figure}[!t]
\centering
\includegraphics[width=\linewidth]{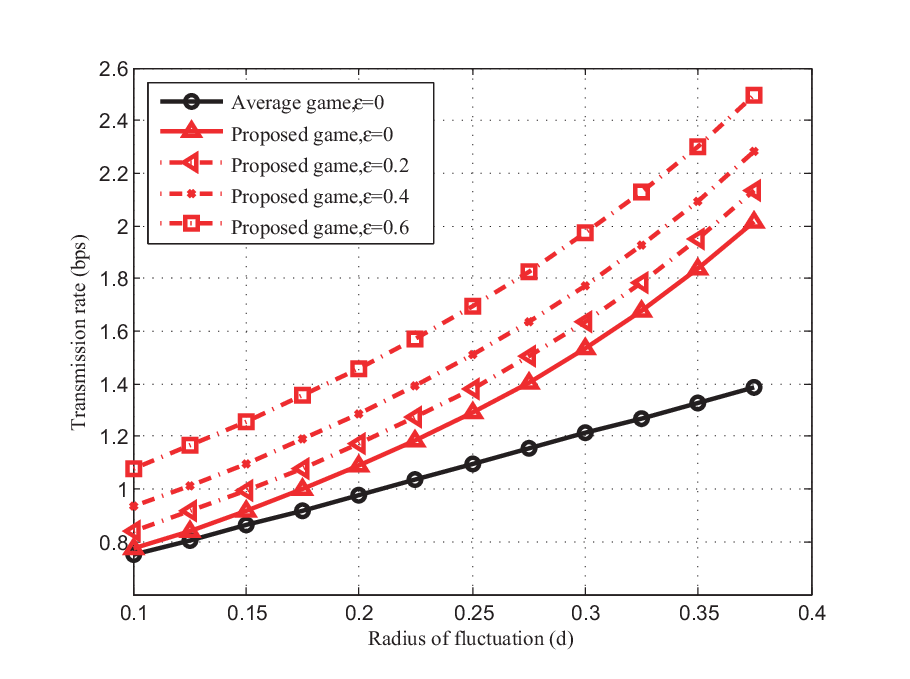}
\caption{The performance comparison of different solutions.}
\label{Fig5}
\end{figure}

In Fig.~\ref{Fig5}, we show the performance comparison of the transmission rate for different solutions. To evaluate the performance, we compare the proposed Bayesian anti-jamming Stackelberg game with the average game, which is a modified game in \cite{existwork6} and only concerns the average value of the incomplete information, e.g., channel state information and transmission cost. It can be seen that the transmission rate of the proposed game is better than the average game. In addition, the observation error $\varepsilon$ results in the increase of the user's transmission rate.

\subsection{Anti-jamming channel selection game in dense wireless networks}

In this section, we investigate the anti-jamming channel selection problem in the dense wireless networks with time-varying radio environment, where the mutual co-channel interference among users and external malicious jamming exist simultaneously. On the one hand, the external malicious jamming is a significant factor to deteriorate the system performance of the wireless networks. One the other hand, the mutual co-channel interference also significantly degrades the system performance. In \cite{existwork14}, a system with \emph{N} users and one jammer is considered, and a single leader and multi-follower Stackelberg game is formulated. To obtain the desirable solutions, we propose a hierarchical learning algorithm (HLA), which can cope with the uncertain, dynamic and incomplete constraints. The implementation process is shown in Fig.~\ref{process}(b). To be specific, for the jammer, based on \emph{Q}-learning, a channel selection algorithm is proposed, and the jammer's strategies are updated at each epoch \emph{k}. For users, a channel selection algorithm based on stochastic learning automata (SLA) is proposed, and the users' strategies update at each time slot \emph{t}. It is noted that each epoch contains \emph{K} time slots. The metric is the expected weighted aggregate interference and jamming (EWAIJ), and it is defined as:
 \begin{equation}
\begin{array}{c}
U = \sum\limits_{n \in {\cal N}} {\sum\limits_{m \in \left\{ {{\cal N}/\left\{ n \right\}} \right\}} {{P_n}{P_m}\bar H_{mn}^{{a_n}}} f({a_m},{a_n})} \\
 + \sum\limits_{n \in {\cal N}} {{P_n}{P_j}\bar H_{jn}^{{c_j}}f({c_j},{a_n}),}
\end{array}
\end{equation}
 where ${a_n}$ denotes the channel selection of user \emph{n}, ${c_j}$ denotes the jamming channel, ${\cal N}$ is the user set, $\bar H_{mn}^{{a_n}}$ represents the expected interference gain from user \emph{m} to user \emph{n}, $\bar H_{jn}^{{c_j}}$ denotes the expected jamming gain from the jammer to user \emph{n}, $f({a_m},{a_n})$ denotes an indicator function ($f({a_m},{a_n}) = 1$ for ${a_m} = {a_n}$, and $f({a_m},{a_n}) = 0$ for ${a_m} \ne {a_n}$), ${P_m}$ and ${P_j}$ represents the transmission power of user \emph{n} and the jammer, respectively. The rate of user\emph{ n} can be given by:
 \begin{equation}
{R_n} = B\log \left( {1 + \frac{{{P_n} H_{nn}^{{a_n}}}}{{B{N_0} + {I_n} + {J_n}}}} \right),
\end{equation}
 where \emph{B} denotes the bandwidth, ${N_0}$ is the noise power spectrum density, ${I_n}$ and ${J_n}$ represent the co-channel interference and malicious jamming, respectively.

 The utility of user \emph{n} is defined as ${u_n}({a_n},{{\bf{a}}_{ - n}},{c_j}) = L - \sum\nolimits_{m \in \left\{ {{\cal N}/\left\{ n \right\}} \right\}} {{P_n}{P_m}\bar H_{mn}^{{a_n}}f({a_m},{a_n})}  - {P_n}{P_j}\bar H_{jn}^{{c_j}}f({c_j},{a_n})$, where \emph{L} represents a predefined positive constant. Each user aims to maximize its utility. The jammer's utility is ${u_j}\left( {{\bf{a}},{c_j}} \right) = \sum\nolimits_{n \in {\cal N}} {{P_n}{P_j}\bar H_{jn}^{{c_j}}f({c_j},{a_n})}$, and it aims at maximizing its damage.
Our objective is to find a channel selection profile that can minimize EWAIJ, which denotes the received mutual interference and external malicious jamming.

\begin{figure*}[!t]
\subfigure[Comparison results of the EWAIJ (${P_j} = 25W$).]{
\label{fig:mini:subfig:a} %% label for first subfigure
\begin{minipage}[b]{0.32\textwidth}
\centering
\includegraphics[width=\linewidth]{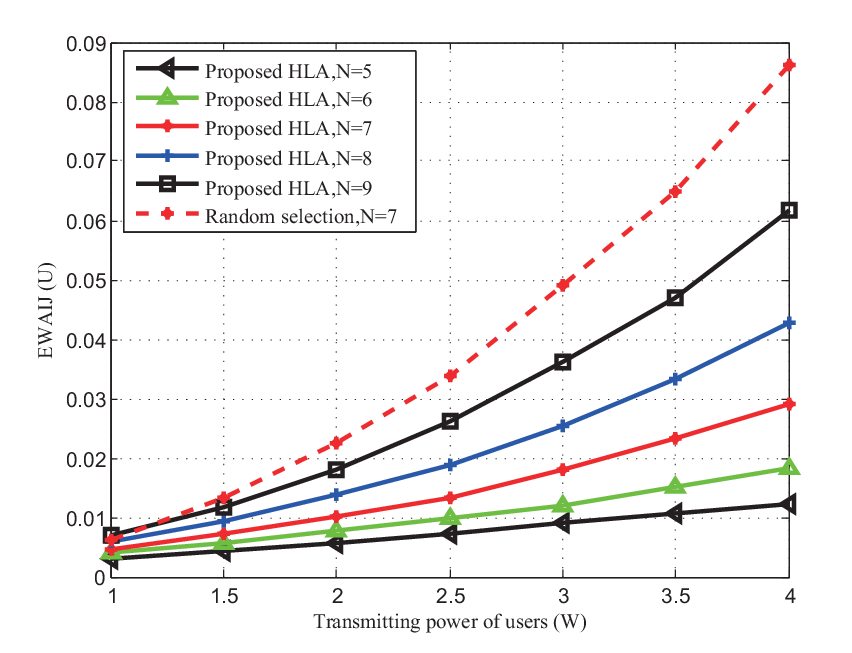}
\end{minipage}}%
\subfigure[Comparison results of the expected achievable rate (${P_j} = 25W$).]{
\label{fig:mini:subfig:b} %% label for second subfigure
\begin{minipage}[b]{0.32\textwidth}
\centering
\includegraphics[width=\linewidth]{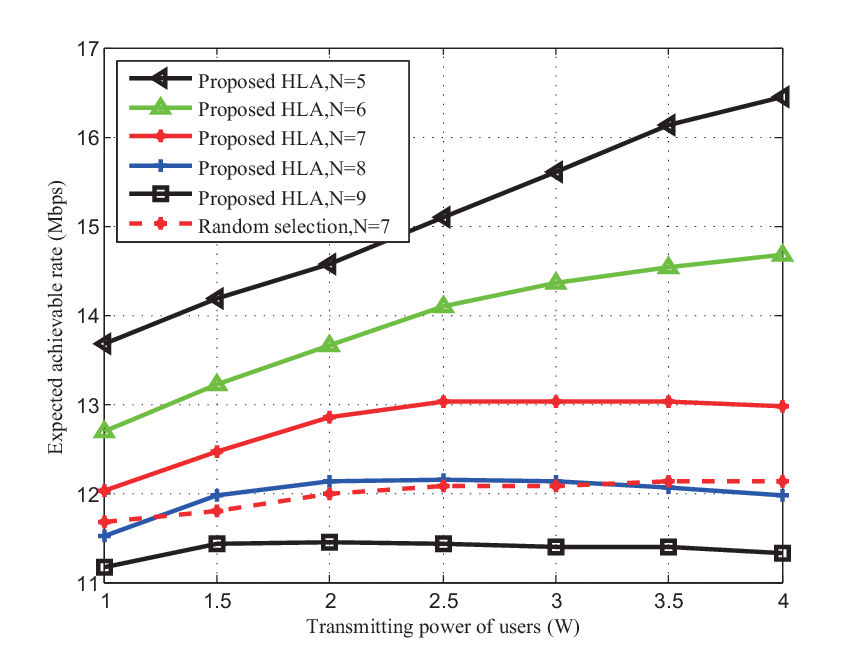}
\end{minipage}}
\subfigure[Comparison results of the expected achievable rate for different solutions (${P_n} = 4W$).]{
\label{fig:mini:subfig:c} %% label for second subfigure
\begin{minipage}[b]{0.32\textwidth}
\centering
\includegraphics[width=\linewidth]{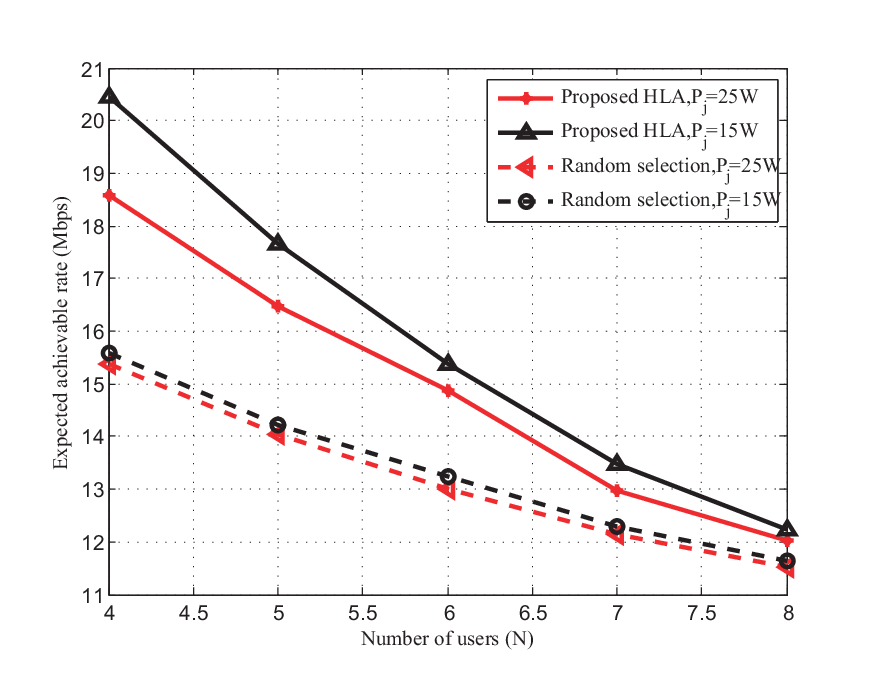}
\end{minipage}}
\caption{The performance comparison.}
\label{performance} %% label for entire figure
\end{figure*}

In Fig.~\ref{performance}, the performance comparison is presented. To show the performance, we compare the proposed HLA algorithm with the random selection scheme, in which the user randomly selects one channel at each time. As indicated in Fig.~\ref{performance}(a), the proposed HLA algorithm outperforms the random selection scheme, and higher transmitting power ${P_i}$ of legitimate users results in the increase of the EWAIJ performance. Moreover, the EWAIJ increases with the growing number of users. The reason is that growing number of users leads to more serious mutual interference. As can be seen from Fig.~\ref{performance}(b), as the number of users increases, the expected achievable rate of each user decreases. The reason is that increasing number of users brings heavy mutual interference. It is noted that the expected achievable rate grows with increasing transmitting power of users at first. Then, if the number of users is large enough, the curve would reach a steady level or even go downwards with higher transmitting power of users. The reason is that higher transmitting power of legitimate users yields more serious mutual interference. As shown in Fig.~\ref{performance}(c), the proposed HLA is superior to the random selection scheme and yields higher expected achievable rate. Specifically, when $N = 4$ and ${P_j} = 15W$, the expected achievable rate of the proposed HLA is improved by approximately 30$\%$ compared to the random selection scheme. It is also noted that the performance gap between the proposed HLA and random selection scheme decreases with the increasing number of users. The reason is that when the number of users \emph{N} becomes sufficiently large, the channels are very crowded, and users can uniformly be spread over the channels. In addition, it can be seen from Fig.~\ref{performance}(c) that the expected achievable rate will decrease with the growing transmitting power of the jammer. The reason is that higher transmitting power of the jammer leads to more serious damage.

%%%%%%%%%%ͨÀ¸Í¼Æ¬ÅÅ°æ%%%%%%
%%%%%%%%%%%%%%%%%%%%%%%%%%%%
%\begin{figure*}[!t]
%\subfigure[Comparison results of the EWAIJ.]{
%\label{fig:mini:subfig:a} %% label for first subfigure
%\begin{minipage}[b]{0.5\textwidth}
%\centering
%\includegraphics[width=\linewidth]{Fig6a}
%\end{minipage}}%
%\subfigure[Comparison results of the expected achievable rate.]{
%\label{fig:mini:subfig:b} %% label for second subfigure
%\begin{minipage}[b]{0.5\textwidth}
%\centering
%\includegraphics[width=\linewidth]{Fig6b}
%\end{minipage}}
%\caption{The performance comparison (${P_j} = 25W$).}
%\label{performance} %% label for entire figure
%\end{figure*}
%%%%%%%%%%ͨÀ¸Í¼Æ¬ÅÅ°æ½áÊø%%%%%%

%%%%%%%%%ͨÀ¸Í¼Æ¬ÅÅ°æ%%%%%%
%%%%%%%%%%%%%%%%%%%%%%%%%%%
%%%%%%%%%ͨÀ¸Í¼Æ¬ÅÅ°æ½áÊø%%%%%%

%\begin{figure}[!t]
%\centering
%\includegraphics[width=\linewidth]{WINE_EWAIJ}
%\caption{The performance comparison (${P_j} = 25W$).}
%\label{WINE_EWAIJ}
%\end{figure}

 \section{Future research directions}
%%%%%%%%%%%%%%%%%%%%%%%%
Although the study to the anti-jamming schemes based on Stackelberg game is now in infancy, it will definitely attract great interest and attention in the near future. Based on the previous discussions, we give some future research directions in the following.

\begin{itemize}
\item To improve the jamming efficiency and decrease the probability of detection, the position of the jammer may be mobile in practical scenarios. Therefore, it is interesting to investigate the anti-jamming problems in mobile scenarios.
\item In realistic scenarios, the players' utility functions are not deterministic and random due to variable channel state information, noise, feedback errors, and observation error. It is, therefore, interesting to study the anti-jamming problem with estimation error. For this scenario, a noisy stochastic game can be formulated, and efficient estimation samples can be used to improve the anti-jamming performance.
\item In dense wireless networks, we need to jointly consider mutual interference among legitimate users and external malicious jamming. Moreover, various traffic demands and priorities also need to be investigated. Specifically, different traffic demands will lead to variable set of active users, and it results in variable set of players. Thus, a dynamic game model with dynamic traffic demands is needed. Different priorities represent the degree of importance, and the users with higher priority should be satisfied with their requirements at first. Consequently, it is promising to investigate the anti-jamming schemes with dynamic traffic demands and different priorities in wireless networks.
\item Considering the heterogeneity in future wireless networks, the user devices may have various service demands and different anti-jamming requirements. Moreover, in a heterogeneous wireless network, intra-system interference, inter-system interference, and external malicious jamming co-exist, and therefore, constitute a complicated environment. Thus, it is a challenging task to design the anti-jamming approaches in heterogeneous wireless networks.
\item According to the history observations of the external jammers, we can predict the action of the jammers, and further enhance the anti-jamming performance. Therefore, it is an interesting topic to design effective anti-jamming schemes with jamming prediction ability in future work.
\end{itemize}

\section{Conclusion}
In this article, we investigated the anti-jamming defence problem in wireless communications, which is an important topic in wireless transmission, and also a vital problem for the security of the spectrum availability. First, we presented the technical challenges and fundamental requirements of anti-jamming defence problem in wireless networks. Then, following the advantages of the Stackelberg game model in anti-jamming field, an anti-jamming decision-making framework based on Stackelberg game was established. We provided two case studies to analyze the anti-jamming Stackelberg game. Finally, we gave some future research directions.

%\section{Acknowledgements}
%This work was supported in part by the Natural Science Foundation for Distinguished Young Scholars of Jiangsu Province under Grant BK20160034, in part by the National Science Foundation of China under Grant 61631020, Grant 61671473, and Grant 61771488, and in part by the Open Research Foundation of Science and Technology in Communication Networks Laboratory.

% Note that IEEE typically puts floats only at the top, even when this
% results in a large percentage of a column being occupied by floats.


\ifCLASSOPTIONcaptionsoff
  \newpage
\fi

\end{document}